\documentclass[%
reprint,
 amsmath,amssymb,
 aps, physrev,
pra,
]{revtex4-2}

\usepackage{graphicx}
\usepackage{dcolumn}
\usepackage{bm}
\usepackage{subcaption}
\usepackage{physics}

\usepackage[percent]{overpic} 

\begin{document}

\title{Double-slit optical ventriloquism: High phase sensitivity via diffraction patterns}

\author{Merav Kahn}
\email{merav.kahn@mail.huji.ac.il} 
\affiliation{Racah Institute of Physics, The Hebrew University of Jerusalem, Jerusalem 91904, Israel}

\author{John C. Howell}
\affiliation{Institute for Quantum Studies, Chapman University, 1 University Drive, Orange, California 92866, USA}
\affiliation{Racah Institute of Physics, The Hebrew University of Jerusalem, Jerusalem 91904, Israel}
\author{Nadav Katz}
\affiliation{Racah Institute of Physics, The Hebrew University of Jerusalem, Jerusalem 91904, Israel}

\date{\today}

\begin{abstract}
High-sensitivity phase sensing is traditionally performed using complex interferometric configurations. As an alternative, we present a robust and simple system based on the classical Young's double-slit experiment that leverages the "optical ventriloquism" effect to amplify phase signals. This phenomenon arises from super-oscillations near intensity minima, which cause an anomalous shift of the local wave vector as a consequence of weak-value behavior. In this work, we constructed an accessible experimental setup that translates minute phase differences into measurable spatial displacements of the diffraction pattern. We compare the detection performance of an sCMOS camera and a quadrant-cell detector, analyzing the noise sources that limit the system's sensitivity. Our results demonstrate that engineering the dark regions of simple diffraction patterns can provide a foundation for advanced optical sensing technologies with minimal structural complexity.
\end{abstract}

\maketitle

\section{Introduction}

Standard optical phase sensing typically relies on precision interferometry\cite{ monnier2007phases, dalidet2025accurate, zemlyanov2020phase, ataman2018phase}. While powerful, traditional interferometers often require complex configurations with multiple optical components to split, delay, and recombine coherent beams \cite{huang2024review, lashari2023recent}. Such setups demand meticulous alignment and are frequently sensitive to mechanical instabilities.

In this work, we demonstrate a high-sensitivity phase-sensing platform based on the most fundamental configuration in optics: the classical Young’s double-slit experiment \cite{YoungITB}. The primary advantage of this approach lies in its extraordinary structural simplicity. Unlike conventional interferometers that require intricate recombination optics, our system extracts phase information directly from the self-interfering diffraction pattern.

The core of our method lies in operating within the regime of Weak Values \cite{WeakValue, berry2010typical, hosoya2010strange, pusey2014anomalous} by leveraging the Optical Ventriloquism effect \cite{OpticalVentriloquism}. This phenomenon occurs near the intensity minima of the diffraction pattern, where the interfering waves nearly cancel each other out. In these regions, the field exhibits super-oscillations \cite{aharonov1990,superoscillations,superoscillationsBerry1994Faster}, which are local oscillations much faster than the highest frequency component of the light's momentum spectrum \cite{superoscillationsBerry1994Faster, berry2006evolution,katzav2013yield, dennis2008superoscillation,zheludev2022optical}. Consequently, the local wave vector ($k_{loc}$), which represents the phase gradient, undergoes a dramatic amplification and ``bends'' significantly far beyond its nominal range.

This anomalous behavior means that a minute shift in the relative phase between the slits is translated into a large change in the local wave vector's direction. To a detector, it appears as if the light originates from a fictitious source spatially displaced from the actual slits, hence the term ventriloquism. By detecting these amplified transverse displacements of the intensity centroid, we transform a simple pedagogical setup into a robust, high-precision metrological tool.

To evaluate the practical capabilities of this approach, we implement an experimental setup where phase shifts are induced by varying the optical path length through a glass medium in one of the slits. We perform a systematic comparative analysis between two distinct detection strategies: a pixel-array sCMOS camera for detailed spatial profiling and a quadrant-cell photodetector for high-speed tracking. By analyzing the fundamental noise limits and calibration factors of each system, we demonstrate that this simple diffractive architecture can approach the shot-noise limit, offering a versatile and easy-to-implement framework for precision phase metrology.

\section{Theoretical Background}

The physical basis of the optical ventriloquism effect \cite{OpticalVentriloquism} utilized in this work lies in the propagation of a monochromatic scalar wave, $\Psi(\mathbf{r}, t) = \exp(-i\omega t)\psi(\mathbf{r})$, where the spatial amplitude is expressed in polar form as 
\begin{equation}
    \psi(\mathbf{r}) = A(\mathbf{r})\exp[ik_0\phi(\mathbf{r})].
\end{equation} 

Within this framework, the local wave vector $\mathbf{k}_{loc}(\mathbf{r})$ is determined by the phase gradient:

\begin{equation} \label{WeakValue}
\mathbf{k}_{loc}(\mathbf{r}) = k_0\nabla\phi = \text{Re} \left[ \frac{\langle \mathbf{r} | \hat{\mathbf{k}} | \psi \rangle}{\langle \mathbf{r} | \psi \rangle} \right]
\end{equation}

By identifying the field as a state vector $\psi(\mathbf{r}) \equiv \langle \mathbf{r} | \psi \rangle$, the local wave vector can be interpreted as the \textit{weak value} \cite{WeakValue} of the wave-vector operator $\hat{\mathbf{k}}$. In this expression, the denominator represents the post-selection of the field at a specific spatial location $\langle \mathbf{r} |$. The wave-vector operator $\hat{\mathbf{k}}$ is defined in a manner analogous to the quantum momentum operator, such that its projection onto the spatial basis follows the relation: $\langle \mathbf{r} | \hat{\mathbf{k}} | \psi \rangle = -i \nabla \langle \mathbf{r} | \psi \rangle$.

In our double-slit configuration, phase information is extracted through such post-selection in regions of destructive interference. In these intensity minima, the local wave vector exhibits anomalous values, known as \textit{super-oscillations} \cite{superoscillations}, that far exceed the momentum spectrum of the source. This phenomenon arises where standard geometric optics fails, as described by the extended eikonal equation derived from the Helmholtz equation: 

\begin{equation}
|\mathbf{k}_{loc}|^2 = (n\omega/c)^2 + \frac{\nabla^2 A}{A}
\end{equation}

Near a diffraction minimum, the amplitude $A$ vanishes rapidly, causing the ``quantum potential'' term $\nabla^2 A / A$ to dominate and induce a self-bending effect on the light rays, a phenomenon termed optical ventriloquism. Consequently, the field appears to originate from a fictitious source spatially displaced from the actual source.

The interference pattern of the two beams is governed by the difference between their transverse wavevectors, defined as $\Delta k \equiv k_{2x} - k_{1x}$. Here, $k_{1x}$ and $k_{2x}$ are the transverse wavevector components of the beams emerging from each slit. These components differ due to a phase manipulation performed on one of the slits, specifically the introduction of a glass slide of a certain thickness, which alters the optical path length relative to the air-filled path of the other slit. The cornerstone of this metrological approach is the mathematical identity:

\begin{equation}
\frac{\partial k_{loc}}{\partial (\Delta\phi)} = \frac{1}{\Delta k} \frac{\partial^2 \phi}{\partial x^2}
\end{equation}

where $\partial^2 \phi / \partial x^2$ is the local phase curvature and $\Delta\phi$ is the relative phase shift between the slits. At intensity minima, this curvature is maximized, leading to \textit{phase-gradient amplification}, where an infinitesimal variation in the relative phase $\Delta\phi$ is translated into a significant shift of the local wave vector $\mathbf{k}_{loc}$.

This variation in the local wave vector manifests as a transverse displacement of the intensity centroid in the detection plane. By employing a position-sensitive detector, these spatial shifts can be characterized with high precision, enabling the detection of minute phase changes. The ultimate phase sensitivity $\delta\phi$ is governed by the total detected photon number $N_{det}$, calculated as:

\begin{equation}
N_{det} = \frac{P \cdot T \cdot \lambda}{h \cdot c}
\end{equation}

In this expression, $P$ represents the optical power measured at the intensity minimum (the working point), $T$ is the integration time of the measurement, $\lambda$ is the operating wavelength, $h$ is Planck's constant, and $c$ is the speed of light. Based on the interference model, the power at the working point $P_0$ relates to the adjacent maxima by 
\begin{equation} \label{P0Cal}
    P_{max} \approx 4P_0.
\end{equation}
 While the fundamental limit is set by \textit{shot noise} \cite{GerryKnight2005, DemkowiczDobrzanski2015QuantumLimits}
 
\begin{equation}\label{ShotNoiseEq}
\delta\phi = \frac{1}{\sqrt{N_{det}}},
\end{equation}
 
 The use of optical ventriloquism maximizes the information extracted from the intensity minima, allowing the system to approach these fundamental metrological bounds.

\section{Experimental Setup}

\begin{figure}[t]
     \centering
     \begin{subfigure}[b]{0.48\textwidth}
         \centering
         \includegraphics[width=\textwidth]{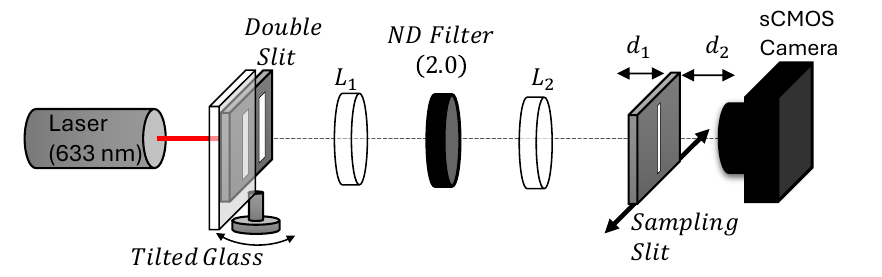}
         \caption{Camera-based detection setup.}
         \label{set_up_camera}
     \end{subfigure}
     
     \vspace{1cm} 
     
     \begin{subfigure}[b]{0.48\textwidth}
         \centering
         \includegraphics[width=\textwidth]{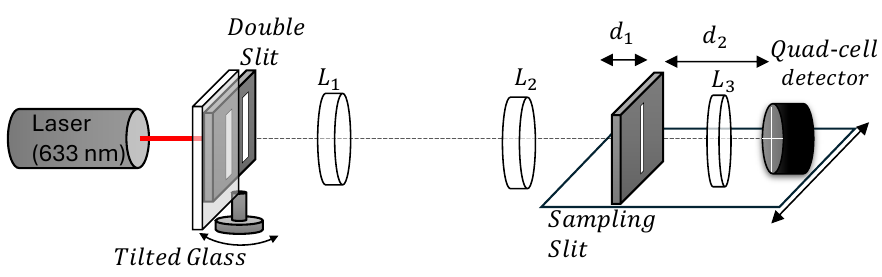}
         \caption{Quad-cell detection setup.}
         \label{set_up_Quad-cell}
     \end{subfigure}
     
     \caption{Experimental configurations for phase sensing via optical ventriloquism. Both setups share a common interferometric backbone: a 633-nm He-Ne laser illuminates a double-slit aperture, where a tilted glass plate introduces a relative phase shift $\Delta\phi$. The resulting interference pattern is relayed through a $4f$ lens system ($L_1, L_2$) to a sampling slit positioned at a distance $d_1$. (a) Camera-based configuration: A stationary sCMOS camera, located $d_2$ behind the moving sampling slit, records the intensity profile. An ND filter is used to prevent detector saturation. (b) Quad-cell configuration: The sampling slit and the quad-cell detector are co-mounted on a translation stage at a fixed separation $d_2$. A focusing lens $L_3$  is positioned between them to concentrate the light onto the detector's active area for high-speed position sensing.}
     \label{fig:full_setup}
\end{figure}

To implement high-sensitivity phase sensing via optical ventriloquism, we developed two experimental configurations sharing a common optical backbone, as illustrated in Figs.~\ref{set_up_camera} and~\ref{set_up_Quad-cell}. In both setups, a 633-nm He-Ne laser beam is incident on a double-slit aperture with a slit width of $w = 100\,\mu\text{m}$ and a center-to-center separation of $s = 600\,\mu\text{m}$. A 1-mm-thick glass plate, mounted on a motorized precision rotation mount (Thorlabs PRM1Z8), is positioned to cover only one of the slits. By adjusting the rotation angle of the plate, we introduce a controlled relative phase shift, $\Delta\phi$, between the two interfering beams.
The diffracted light is relayed through a $4f$ imaging system consisting of two lenses with focal lengths of $f = 100\,\text{mm}$. To probe the diffraction field in the Fresnel region, a sampling slit ($w = 5\,\mu\text{m}$, Thorlabs S5K) is positioned at a distance $d_1 \approx 3.7\,\text{cm}$ beyond the $4f$ image plane. The sampling slit is mounted on a linear translation stage driven by a compact stepper motor actuator (Thorlabs ZFS25B), enabling a precise spatial scan of the interference pattern.

In the first configuration [Fig.~\ref{set_up_camera}], a neutral density (ND) filter (OD 2.0) is placed between the $4f$ lenses to attenuate the beam and prevent detector saturation. A monochrome sCMOS camera (Thorlabs CS2100M-USB) is positioned behind the scanning slit to record the intensity distribution. The local beam position at each stage coordinate is determined by calculating the intensity centroid (the mean $x$-position) of the recorded profile.

In the second configuration [Fig.~\ref{set_up_Quad-cell}], the ND filter is removed to maintain sufficient signal for the quadrant detector. Here, the sampling slit and a quad-cell detector (New Focus Photoreceivers, Model 2921) are co-mounted on the translation stage. A lens with a focal length of $f = 25.4\,\text{mm}$ is placed between them to focus the diffracted order from the slit onto the detector's active area. The transverse beam position is derived from the normalized difference signal, calculated as the ratio of the difference output ($X_{\text{diff}}$) to the sum output ($V_{\text{sum}}$). This normalization ensures the position measurement is independent of total power fluctuations. In both configurations, we record both the local power and the beam displacement as a function of the stage coordinates.

\section{Results and Discussion}



\begin{figure}[h]
\centering
\captionsetup[subfigure]{labelformat=empty}

\begin{subfigure}[b]{0.48\textwidth}
    \centering
    \begin{overpic}[width=\textwidth]{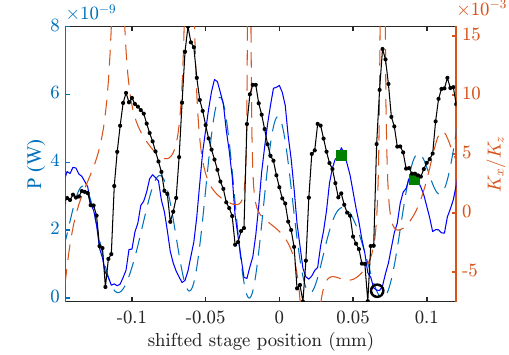}
        \put(14, 59){\textbf{(a)}} 
    \end{overpic}
    \refstepcounter{subfigure}\label{I_Kx_Kz_power_camera}
\end{subfigure}

\vspace{0.01cm} 

\begin{subfigure}[b]{0.46\textwidth}
    \centering
    \begin{overpic}[width=\textwidth]{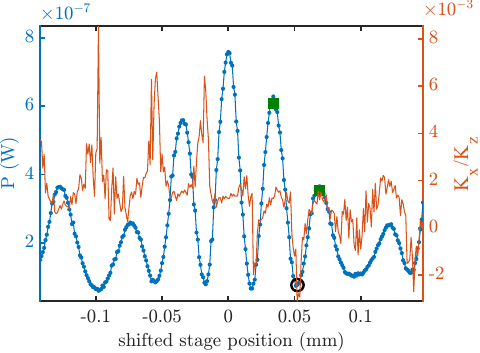}
        \put(10, 63){\textbf{(b)}}
    \end{overpic}
    \refstepcounter{subfigure}\label{I_Kx_Kz_power_scope}
\end{subfigure}

\caption{Spatial characterization of optical power ($P$) and local wave vector ($k_x/k_z$) for both detection modalities: (a) camera-based and (b) quad-cell setups. Results are plotted as a function of the stage position $x$, which is shifted such that the peak power is centered at $x=0$. (a) Camera-based results: dashed and solid-dotted blue lines represent simulated and experimental optical power, respectively. Dashed orange and solid black lines show simulated and experimental $k_x/k_z$ values. (b) Quad-cell results: solid blue and orange lines represent experimental power and $k_x/k_z$ values, respectively. In both panels, the black circle indicates the local minimum near the selected operating point, and the green squares mark the peak positions used to calculate the average power $P_0$.}
\label{fig:power_kx_kz_system}
\end{figure}

The first step of the experiment was to scan the diffraction pattern using the sampling slit in order to identify the optimal operating point, characterized by a sharp peak, in both experimental configurations. To facilitate comparison, the stage position $x$ for both setups was shifted such that the peak power is centered at $x = 0$. Figure~\ref{fig:power_kx_kz_system}(a) presents the scan results for the camera-based setup, including a comparison with the numerical simulation. The  dashed blue line and the solid blue line represent the power profiles, normalized using a power meter, for the simulation and the measurement, respectively. The  dashed orange line and the solid black line display the local wave vector ($k_x/k_z$) for the simulation and the experiment, respectively. As shown, a strong agreement is observed between the experimental data and the theoretical simulation.

A similar characterization was performed for the quad-cell setup, as shown in Fig.~\ref{fig:power_kx_kz_system}(b), where the blue line with points represents the measured optical power and the solid orange line represents the local wave vector.

Based on these spatial profiles, we selected the operating point of the translation stage to be near the peak of the local wave vector, at $x = 0.0664$~mm for the camera setup and $x = 0.0502$~mm for the quad-cell setup. In both cases, a black dot marks the local power minimum near the operating point, and green squares indicate the local power maxima used to determine the average power ($P_0$) for each system according to the relation in Eq. (~\ref{P0Cal}). The average power at the operating point was calculated to be $9.06 \times 10^{-10}$~W for the camera and $1.01 \times 10^{-7}$~W for the quad-cell.



\begin{figure}[h]
\centering
\captionsetup[subfigure]{labelformat=empty}

\begin{subfigure}[b]{0.48\textwidth}
    \centering
    \begin{overpic}[width=\textwidth]{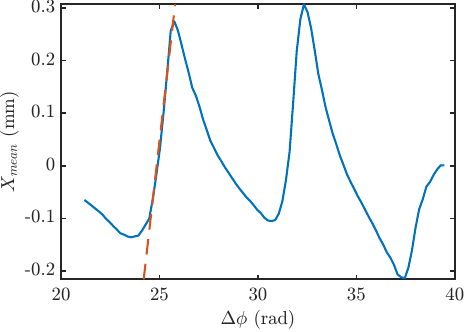}
        \put(15, 65){\textbf{(a)}} 
    \end{overpic}
    \refstepcounter{subfigure}\label{fit_phi_camera}
\end{subfigure}

\vspace{0.5cm}

\begin{subfigure}[b]{0.48\textwidth}
    \centering
    \begin{overpic}[width=\textwidth]{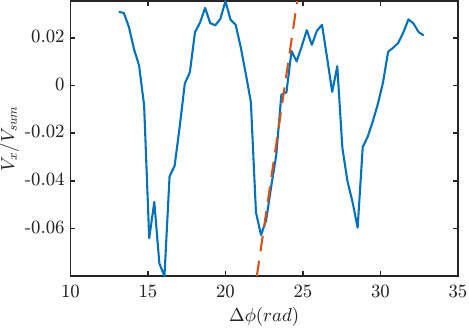}
        \put(16, 65){\textbf{(b)}}
    \end{overpic}
    \refstepcounter{subfigure}\label{fit_phi_scope}
\end{subfigure}

\caption{Calibration of phase-to-displacement sensitivity: (a) camera-based and (b) quad-cell setups. (a) Mean X-position captured by the CMOS camera and (b) normalized horizontal difference signal from the quad-cell detector as a function of the induced phase shift. Solid blue lines represent experimental measurements, while dashed orange lines indicate linear fits at the steepest regions, defining the optimal calibration factors.}
\label{fig:fit_phi_system}
\end{figure}

Following the spatial characterization, the sampling slit was fixed at the previously selected operating point. To calibrate the system, we modulated the relative phase difference between the beam passing through the glass plate and the beam propagating in free space by rotating the glass plate, while measuring the resulting beam displacement. In the camera-based setup, this displacement was determined by calculating the mean X-position of the captured diffraction pattern [Fig.~\ref{fit_phi_camera}]. For the quad-cell setup, the shift was derived from the normalized horizontal difference signal, $V_{\text{x}}/V_{\text{sum}}$ [Fig.~\ref{fit_phi_scope}].

The rotation angle of the glass plate was converted into a phase shift by analyzing the periodicity of the detected intensity. Figure~\ref{fig:fit_phi_system} displays the measured beam shift as a function of the relative phase difference for both detection modalities (solid blue lines). For each system, we identified the region with the steepest slope to maximize phase sensitivity. These linear regions, marked by dashed orange lines in Fig.~\ref{fig:fit_phi_system}, define the calibration factors used to translate experimental spatial shifts into phase measurements.



\begin{figure}[!]
\centering
\captionsetup[subfigure]{labelformat=empty}

\begin{subfigure}{0.48\textwidth}
    \centering
    \begin{overpic}[width=\textwidth]{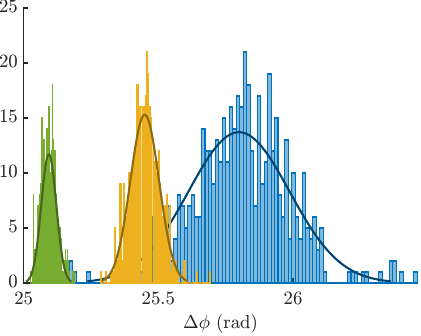}
        \put(10, 73){\textbf{(a)}} 
    \end{overpic}
    \refstepcounter{subfigure}\label{Histogramall_hist_phi_camera}
\end{subfigure}

\vspace{0.8cm} 

\begin{subfigure}{0.48\textwidth}
    \centering
    \begin{overpic}[width=\textwidth]{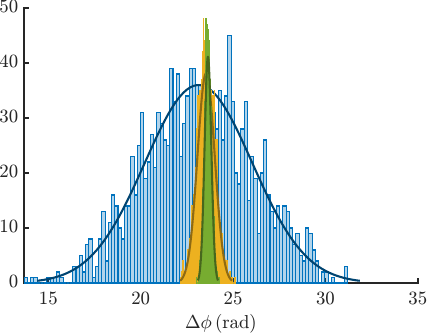}
        \put(10, 70){\textbf{(b)}}
    \end{overpic}
    \refstepcounter{subfigure}\label{Histogramall_hist_phi_scope}
\end{subfigure}

\caption{Relative phase shift distributions for both detection modalities: (a) camera-based and (b) quad-cell setups. Due to the two detectors' disparate sampling capabilities, the distributions are measured across distinct temporal regimes. The blue, yellow, and green histograms denote relative increases in integration time for each setup: for the camera-based setup (a), exposure times were $0.09$, $1$, and $10$~ms, respectively; for the quad-cell setup (b), integration times were $4 \times 10^{-4}$, $10^{-1}$, and $1$~ms, respectively. Solid lines represent Gaussian fits, where the narrowing of the standard deviation ($\sigma_\phi$) quantifies the noise-averaging efficiency.}
\label{fig:Histogramall_hist_phi}
\end{figure}

After determining the calibration factors, we evaluated the system's sensitivity to relative phase shifts at the optimal operating point. To this end, time series of the beam position were recorded under constant phase conditions for various integration times in both experimental setups. The histograms presented in Figs.~\ref{Histogramall_hist_phi_camera} and \ref{Histogramall_hist_phi_scope} display the measured relative phase shift distributions. Each distribution was fitted with a Gaussian function, where the standard deviation ($\sigma_\phi$) represents the total noise level of the system. 

To evaluate the sensitivity limits, we characterized the relative phase shifts across the characteristic temporal regimes of each detector. It is important to emphasize that the integration times for the camera and the quad-cell were selected to span their respective optimal operational scales; consequently, while the color-coding (blue, yellow, green) represents an identical trend of increasing exposure, the absolute time values are setup-specific and reflect the inherent sampling constraints of each modality.

For the camera-based setup [Fig.~\ref{Histogramall_hist_phi_camera}], the measured noise levels (represented by the blue, yellow, and green histograms) were $0.2$, $5.3 \times 10^{-2}$, and $2.8 \times 10^{-2}$~rad for exposure times of $0.09$, $1$, and $10$~ms, respectively. To assess the system's performance relative to the fundamental physical limit, these measured standard deviations were compared to the theoretical shot-noise levels for this setup, which were calculated according to Eq.(~\ref{ShotNoiseEq}) as $1.4 \times 10^{-3}$, $4.1 \times 10^{-4}$, and $1.3 \times 10^{-4}$~rad for the corresponding exposure times.

A similar procedure was performed for the quad-cell setup [Fig.~\ref{Histogramall_hist_phi_scope}], where the measured noise levels were $2.9$, $0.47$, and $0.18$~rad for integration times of $4 \times 10^{-4}$, $10^{-1}$, and $1$~ms, respectively. In comparison to the fundamental limit, the theoretical shot noise for this setup was calculated as $1.97 \times 10^{-3}$, $1.25 \times 10^{-4}$, and $3.95 \times 10^{-5}$~rad.

While increasing the integration time reduces the noise in both setups, the rate of reduction deviates from the ideal $1/\sqrt{T}$ scaling characteristic of white stochastic noise. This behavior indicates that the system's sensitivity is not solely limited by random fluctuations but is dominated by inherent technical and low-frequency environmental noise, such as mechanical vibrations and air currents, which do not average out as efficiently.

The discrepancy between the measured noise and the theoretical shot-noise limit further reveals the specific inherent noise sources of each detection method. In the camera-based setup, the $1920 \times 1080$ pixel array introduces a cumulative noise floor comprised of the independent read noise and dark current from each active pixel. Consequently, the sensitivity is intrinsically tied to the detector's spatial integration; reducing the region of interest (ROI) to minimize the number of pixels involved in the centroid calculation would directly lower this inherent noise contribution. 

In contrast, the quad-cell setup suffers from a different inherent limitation related to the signal normalization process, $V_{\text{diff}}/V_{\text{sum}}$. A residual, uncalibratable electronic offset exists within the detection circuitry, which becomes a critical bottleneck near the intensity minimum. As the denominator $V_{\text{sum}}$ approaches zero, the relative weight of this constant offset is drastically amplified. This inherent error mechanism dominates the signal-to-noise ratio at the operating point, effectively masking the quantum shot noise and preventing the system from reaching its fundamental limit despite prolonged temporal averaging.

\section{Conclusion}

In this work, we have demonstrated a high-sensitivity phase-sensing platform based on the fundamental Young's double-slit experiment, utilizing the phenomenon of optical ventriloquism. We showed that near the intensity minima of the diffraction pattern, where destructive interference occurs, the field exhibits super-oscillations that cause the local wave vector to bend dramatically. This physical mechanism translates minute relative phase shifts between the slits into significant transverse displacements of the beam's intensity centroid.
 
Our experimental investigation characterized this effect using two distinct detection modalities: a pixel-array sCMOS camera and a quadrant-cell detector. The analysis of the noise floor for both systems revealed that the measured sensitivity remained above the theoretical shot-noise limit. In the camera-based setup, performance was constrained by cumulative readout noise and dark current. In contrast, the quadrant-cell setup was primarily limited by an electronic offset that becomes dominant near the intensity minimum as the total signal vanishes. Furthermore, low-frequency environmental noise, such as mechanical vibrations and air currents, was found to limit the ideal integration scaling.

Despite these technical constraints, the primary advantage of this system remains its extraordinary structural simplicity combined with high intrinsic sensitivity. We anticipate that the system's sensitivity can be further enhanced by implementing more stable engineering of the phase-shifting mechanism and designing tailored interference patterns to optimize the intensity at the minima. Such advancements would minimize power loss while maximizing the phase-gradient amplification, leading to an even more robust and sensitive metrological tool that approaches fundamental physical limits.

\begin{acknowledgments}
NK acknowledges support of the ISF-Quantum and EU project OpenSuperQPlus grants.

The authors acknowledge the use of ChatGPT (OpenAI) for assistance with optimization of experimental control scripts, and Gemini (Google) for language polishing and manuscript preparation. These tools were not used to generate scientific content, analyses, or interpretations. All scientific conclusions were developed and verified solely by the authors.
\end{acknowledgments}

\bibliography{apssamp} 

\end{document}